\documentclass[lettersize,journal]{IEEEtran}
\usepackage{amsfonts}
\usepackage{algorithmic}
\usepackage{array}
\usepackage[caption=false,font=normalsize,labelfont=sf,textfont=sf]{subfig}
\usepackage{textcomp}
\usepackage{stfloats}
\usepackage{url}
\usepackage{verbatim}
\usepackage{graphicx}

\usepackage{cite}
\usepackage{amsmath}
\usepackage{amssymb}
\usepackage{tabularx}
\usepackage{color}

\def\BibTeX{{\rm B\kern-.05em{\sc i\kern-.025em b}\kern-.08em
    T\kern-.1667em\lower.7ex\hbox{E}\kern-.125emX}}
\usepackage{balance}

\newcommand{\etal}{\textit{et al.}}

\begin{document}

\title{Conversion Between CT and MRI Images Using Diffusion and Score-Matching Models}

\author{Qing~Lyu,~\IEEEmembership{Member,~IEEE},
        and~Ge~Wang,~\IEEEmembership{Fellow,~IEEE}
    \thanks{All authors are with the Biomedical Imaging Center, Rensselaer Polytechnic Institute, Troy, NY, 12180, USA (email: lyuqing0828@gmail.com, wangg6@rpi.edu).}
    
}

\markboth{Journal of \LaTeX\ Class Files,~Vol.~18, No.~9, September~2020}%
{How to Use the IEEEtran \LaTeX \ Templates}

\maketitle

\begin{abstract}
MRI and CT are most widely used medical imaging modalities. It is often necessary to acquire multi-modality images for diagnosis and treatment such as radiotherapy planning. However, multi-modality imaging is not only costly but also introduces misalignment between MRI and CT images. To address this challenge, computational conversion is a viable approach between MRI and CT images, especially from MRI to CT images. In this paper, we propose to use an emerging deep learning framework called diffusion and score-matching models in this context. Specifically, we adapt denoising diffusion probabilistic and score-matching models, use four different sampling strategies, and compare their performance metrics with that using a convolutional neural network and a generative adversarial network model. Our results show that the diffusion and score-matching models generate better synthetic CT images than the CNN and GAN models. Furthermore, we investigate the uncertainties associated with the diffusion and score-matching networks using the Monte-Carlo method, and improve the results by averaging their Monte-Carlo outputs. Our study suggests that diffusion and score-matching models are powerful to generate high quality images conditioned on an image obtained using a complementary imaging modality, analytically rigorous with clear explainability, and highly competitive with CNNs and GANs for image synthesis.
\end{abstract}


\section{Introduction}
\IEEEPARstart{M}{agnetic} resonance imaging (MRI) and computed tomography (CT) are two most widely used medical imaging modalities. MRI shows soft tissues such as vessels and organs in rich contrast while CT is the method of choice for imaging hard tissues like bones as well as interfaces among air, bone and soft tissues. Due to their complementary characteristics, multi-modality imaging with MRI and CT is often used in clinical practice. For example, radiotherapy requires both CT and MRI, because CT provides an electron density distribution indispensable for treatment planning while MRI outlines tumors and soft tissues~\cite{owrangi2018mri}. However, simultaneous CT-MRI is still a research area, and CT and MRI scans are currently performed in separation, which is not only expensive but also brings about nonrigid misalignment between MRI and CT images~\cite{kearney2020attention}. Developing a simultaneous CT-MRI device may be a solution of this problem, and we have conducted studies to propose top-level designs of such a device~\cite{wang2015vision, peng2022top}. 

Medical image synthesis is a viable approach to solve the aforementioned problem. This approach models a mapping from a given source image to an unknown target image. Conventional image synthesis methods focus on exploiting diverse models, such as dictionary learning and random forest, to extract features predefined by experts~\cite{yu2020medical}. However, these methods are limited to handcrafted feature representations. Recently, deep learning has shown huge potentials and great successes in  medical image analysis tasks, such as denoising~\cite{yang2018low,chen2017low}, super-resolution~\cite{lyu2020mri,hammernik2018learning}, and artifact reduction~\cite{zhang2018convolutional,gjesteby2017deep}. Compared with the traditional methods, deep neural networks learn features in a data-driven fashion and produce superior feature representations. In particular, several deep learning-based cross-modality medical image synthesis studies were reported in recent years, most of which are based on convolutional neural networks (CNNs) and generative adversarial networks (GANs)~\cite{nie2016estimating, han2017mr, leynes2017direct, chartsias2017adversarial, nie2018medical, emami2018generating, hiasa2018cross, zhang2018translating, ben2017virtual, armanious2020medgan, bi2017synthesis, choi2018generation, wei2018learning, cai2019towards, li2020magnetic, bahrami2020new, boni2020mr, ben2019cross, tao2020pseudo, hu2021bidirectional, zhang2022bpgan}.

Diffusion and score-matching models represents an emerging generative approach that has attracted a major attention in the medical imaging field. These models can generate high-fidelity realistic natural images~\cite{song2019generative, ho2020ddpm, song2020score}. Compared with other types of generative models like GANs and variational autoencoders that are difficult to train and interpret, and do not always produce satisfactory image quality, diffusion and score-matching models are analytically principled, and easy to train, and offer state-of-the-art image quality. Impressively, an increasing number of studies show that diffusion and score-matching models beat GANs and variational auto-encoders in multiple image generation tasks~\cite{croitoru2022diffusion}.

In this paper, we propose to use diffusion and score-matching models for image conversion between CT and MRI, with an emphasis on mapping from MRI to CT images. Two models are based on in our study, the denoising diffusion probabilistic model (DDPM)~\cite{ho2020ddpm} and the model solving the stochastic differential equation (SDE)~\cite{song2020score}. These models are compared with CNN and GAN models of consistent network architectures. With the diffusion models, we further quantify their uncertainties from Monte-Carlo sampling results and obtain superior results by averaging these random samples.

\section{Related works}
\subsection{Deep medical image synthesis}
Inspired by the success of deep learning in the computational vision domain, researchers applied deep learning in medical imaging tasks. For  medical image synthesis, CNN and GAN models were proposed. For image conversion between MRI and CT modalities, Nie~\etal~\cite{nie2016estimating} proposed a 3D fully convolutional network to synthesize pelvic CT images from the corresponding MRI images. Then, they proposed a cascaded GAN model with multiple sequential generators and discriminators~\cite{nie2018medical}. Bahrami~\etal~\cite{bahrami2020new} designed a CNN model based on an encoder-decoder backbone and reported that the proposed model exhibited a fast convergence rate with a low number of training subjects. Han~\etal~\cite{han2017mr} and Leynes~\etal~\cite{leynes2017direct} both synthesized CT images using UNet. GAN models with ResNet were created by Emami~\etal~\cite{emami2018generating} and Tao~\etal~\cite{tao2020pseudo}. Boni~\etal~\cite{boni2020mr} built a conditional GAN to generate synthetic CT images based on multi-center pelvic datasets. Furthermore, Chartsias~\etal~\cite{chartsias2017adversarial}, Hiasa~\etal~\cite{hiasa2018cross}, Zhang~\etal~\cite{zhang2018translating}, and Cai~\etal~\cite{cai2019towards} adopted CycleGAN to generate synthetic MRI and CT images. Li~\etal~\cite{li2020magnetic} compared the performance of UNet, cycleGAN, and pix2pix models on converting brain CT scans into MRI images, and found that UNet had the best performance among all the three models.

For image synthesis between CT and positron emission tomography (PET) modalities, Ben-Cohen~\etal~\cite{ben2017virtual} proposed a conditional GAN with a fully convolutional network for liver PET image synthesis. Armanious~\etal~\cite{armanious2020medgan} built a model called MedGAN with cascaded encoder-decoders to minimize a complicated objective function. Bi~\etal~\cite{bi2017synthesis} designed a multi-channel GAN model with a ability to represent semantic information. Ben-Cohen~\etal~\cite{ben2019cross} used a GAN model with a fully connected network to generate synthetic PET images and improve lesion detection.

As far as image synthesis between MRI and PET modalities is concerned, Wei~\etal~\cite{wei2018learning} proposed a sketcher-refiner scheme with two cascaded GANs. The first GAN generates coarse synthetic images. The second GAN refines the results. Choi~\etal~\cite{choi2018generation} built a GAN model with UNet as the generator for MRI image synthesis. Zhang~\etal~\cite{zhang2022bpgan} proposed a model called BPGAN to synthesize brain PET images. Hu~\etal~\cite{hu2021bidirectional} designed a 3D end-to-end synthesis model called the bidirectional mapping generative adversarial network (BMGAN), in which the image context and the latent vector were jointly optimized for brain MRI-to-PET image synthesis.

\subsection{Diffusion and score-matching model}
Diffusion and score-matching models are emerging as most promising deep generative models, with impressive generative capabilities for many tasks such as image generation, super-resolution, and image inpainting~\cite{croitoru2022diffusion}. Such a model usually consists of two stages: a forward stage to gradually add noise, and a reverse stage to denoise and recover an original sample step-by-step. Currently, representative frameworks in this category of image generation methods include denoising diffusion probabilistic models (DDPM)~\cite{ho2020ddpm}, noise conditioned score networks (NCSN)~\cite{song2019generative}, and stochastic differential equations (SDE)~\cite{song2020score}.

\subsubsection{Denoising diffusion probabilistic models}
DDPM has its diffusion stage including multiple small steps. In each step, a data sample such as an image is slightly corrupted by Gaussian noise. Let $x_0$ represents an original image and $q(x_0)$ denotes the original distribution of $x_{0}$, we have $x_{0}\sim q(x_{0})$. A sequence of gradually corrupted images $x_{1}, x_{2}, \ldots , x_{T}$ after each diffusion step can be computed as the following Markovian process: 
\begin{equation}\label{equ_1}
q(x_{t}\vert x_{t-1}) = \mathcal{N}(x_{t}; \sqrt{1-\beta_{t}}\cdot x_{t-1}, \beta_{t}\cdot \mathbf{I}), 
\end{equation}
\begin{equation}\label{equ_2}
q(x_{1:T}\vert x_{0}) = \prod_{t=1}^{T} q(x_{t}\vert x_{t-1}), 
\end{equation}
where $T$ is the total number of noising steps, $\beta_{t} \in (0, 1), $ is a hyper-parameter controlling the variance of incremental Gaussian noise, and $\mathcal{N}(x; \mu, \sigma)$ represents a Gaussian distribution of mean $\mu$ and covariance $\sigma$. With the parametrization $\alpha_{t}=1-\beta_{t}$ and $\bar{\alpha}_{t} = \prod_{i=1}^{t}\alpha_{i}$, we have 
\begin{equation}\label{equ_3}
q(x_{t}\vert x_{0}) = \mathcal{N}(x_{t}; \sqrt{\bar{\alpha}_{t}}\cdot x_{0}, (1-\bar{\alpha}_{t})\cdot \mathbf{I}).
\end{equation}
When $T \rightarrow \infty$, $x_{T}$ becomes an isotropic Gaussian distribution.

In its reverse stage, DDPM performs a denoising task to recover an original image. According to~\cite{sohl2015deep}, each step $q(x_{t-1} \vert x_{t})$ is also a Gaussian distribution if $\beta_{t}$ is small. Then, we can train a neural network $p_{\theta}$ to approximate each reserve diffusion step and estimate the mean $\mu_{\theta}(x_{t}, t)$ and the covariance $\Sigma_{\theta}(x_{t}, t)$:
\begin{equation}\label{equ_4}
p_{\theta}(x_{0:T}) = p(x_{T}) \prod_{t=1}^{T} p_{\theta}(x_{t-1} \vert x_{t}),
\end{equation}
\begin{equation}\label{equ_5}
p_{\theta}(x_{t-1} \vert x_{t}) = \mathcal{N}(x_{t-1}; \mu_{\theta}(x_{t}, t), \Sigma_{\theta}(x_{t}, t)),
\end{equation}
where $p(x_{T})$ is the density function of $x_{T}$. According to~\cite{ho2020ddpm}, the reverse step is tractable conditioned on $x_{t}$ and $x_{0}$:
\begin{equation}\label{equ_6}
q(x_{t-1} \vert x_{t}, x_{0}) = \mathcal{N}(x_{t-1}; \tilde{\mu}(x_{t}, x_{0}), \tilde{\beta}_{t}\cdot \mathbf{I}),
\end{equation}
\begin{equation}\label{equ_7}
\tilde{\mu}(x_{t}, x_{0}) = \frac{\sqrt{\bar{\alpha}_{t-1}}\beta_{t}}{1-\bar{\alpha}_{t}}x_{0} + \frac{\sqrt{\alpha_{t}}(1-\bar{\alpha}_{t-1})}{1-\bar{\alpha}_{t}}x_{t},
\end{equation}
\begin{equation}\label{equ_8}
\tilde{\beta}_{t} = \frac{1-\bar{\alpha}_{t-1}}{1-\bar{\alpha}_{t}}\beta_{t}.
\end{equation}
Given that $x_{0} = \frac{1}{\sqrt{\bar{\alpha}_{t}}}(x_{t} - \sqrt{1-\bar{\alpha}_{t}}\epsilon_{t})$, where $\epsilon_{t} \sim \mathcal{N}(0, \mathbf{I})$. Equation~\eqref{equ_7} can be rewritten as 
\begin{equation}\label{equ_9}
\begin{split}
    \mu_{\theta}(x_{t}, t) &= \tilde{\mu}(x_{t}, \frac{1}{\sqrt{\bar{\alpha}_{t}}}(x_{t} - \sqrt{1-\bar{\alpha}_{t}}\epsilon_{\theta}(x_{t}))) \\
    &= \frac{1}{\sqrt{\alpha_{t}}}(x_{t} - \frac{1-\alpha_{t}}{\sqrt{1-\bar{\alpha}_{t}}}\epsilon_{\theta}(x_{t}, t)),
\end{split}
\end{equation}

The objective for training the noise estimation network $\epsilon_{\theta}$ (added noise $\epsilon_{t}$ in $x_{t}$) is to optimize the variational lower bound (VLB):
\begin{equation}\label{equ_10}
\mathcal{L}_{VLB} = \mathcal{L}_{0} + \sum_{t=1}^{T-1}\mathcal{L}_{t} + \mathcal{L}_{T},
\end{equation}
\begin{equation}\label{equ_11}
\mathcal{L}_{0} = -\log p_{\theta}(x_{0} \vert x_{1}),
\end{equation}
\begin{equation}\label{equ_12}
\mathcal{L}_{t} = KL(q(x_{t-1} \vert x_{t}, x_{0}) \| p_{\theta}(x_{t-1} \vert x_{t})),
\end{equation}
\begin{equation}\label{equ_13}
\mathcal{L}_{T} = KL(q(x_{T} \vert x_{0}) \| p_{\theta}(x_{T})),
\end{equation}
where $KL$ denotes the Kullback-Leibler divergence between two probability distributions. Note that $\mathcal{L}_{T}$ is constant and can be ignored because $q(x_{T} \vert x_{0})$ has no learnable parameters and $x_{T}$ is a Gaussian noise. 

In~\cite{ho2020ddpm}, $\mathcal{L}_{0}$ is computed from $\mathcal{N}(x_{0}; \mu_{\theta}(x_{1}, 1), \Sigma_{\theta}(x_{1}, 1))$, and the loss term in~\eqref{equ_12} can be reparameterized and simplified as
\begin{equation}\label{equ_14}
\begin{split}
    \mathcal{L}_{t}^{simple} = &\mathbb{E}_{t \sim [1,T],x_{0},\epsilon_{t}} [\| \epsilon_{t} - \epsilon_{\theta}(\sqrt{\bar{\alpha}_{t}}x_{0} \\
    &+ \sqrt{1-\bar{\alpha}_{t}}\epsilon_{t}, t) \|^{2}].
\end{split}
\end{equation}
The final simplified objective function is 
\begin{equation}\label{equ_15}
\mathcal{L}_{simple} = \mathcal{L}_{t}^{simple} + C,
\end{equation}
where $C$ is a constant independent of the vector of parameters $\theta$.

\subsubsection{Noise conditioned score network}
Langevin dynamics produces samples from a probability density function $p(x)$ only using the score function $\nabla_{x} \log p_{t}(x)$. Given a fixed step size $\gamma > 0$ and an initial value $x_{0} \sim \pi(x)$ with $\pi$ being a prior distribution, the sampling process using the Langevin method can be expressed as
\begin{equation}\label{equ_16}
x_{t} = x_{t-1} + \frac{\gamma}{2}\nabla_{x} \log p(x) + \sqrt{\gamma} \cdot \omega_{t}, \forall t \in \{1, \ldots, T\},
\end{equation}
where $\omega_{t} \in \mathcal{N}(0, \mathbf{I})$. The distribution of $x_{T}$ equals $p(x)$ when $\gamma \rightarrow 0$ and $T \rightarrow \infty$. A neural network $s_{\theta}$ is trained to estimate the score so that $s_{\theta}(x, t) \approx \nabla_{x} \log p_{t}(x)$. Ideally, the network can be trained via score matching based on the following objective function
\begin{equation}\label{equ_17}
\mathcal{L}_{sm} = \mathbb{E}_{x\sim p(x)} \| s_{\theta}(x) - \nabla_{x} \log p(x) \|_{2}^{2}.
\end{equation}
However, equation~\eqref{equ_17} is hard to be optimized as the score $\nabla_{x} \log p(x)$ is not easy to obtain. To overcome this difficulty, Song~\etal~\cite{song2019generative} proposed to perturb the original data distribution by Gaussian noises at different scales: $\sigma_{1} < \sigma_{2} < \cdots < \sigma_{T}$ such that $p_{\sigma_{1}}(x) \approx p(x_{0})$ and $p_{\sigma_{T}}(x) \approx \mathcal{N} (0, \mathbf{I})$. Then, a NCSN is trained for score estimation: $s_{\theta}(x, \sigma_{t}) \approx \nabla_{x} \log p_{\sigma_{t}}(x), \forall t \in \{1, \ldots, T\}$. Then, we have
\begin{equation}\label{equ_18}
\nabla_{x_{t}} \log p_{\sigma_{t}}(x_{t} \vert x) = -\frac{x_{t}-x}{\sigma_{t}}.
\end{equation}
Combining~\eqref{equ_17} and~\eqref{equ_18} on all $(\sigma_{t})_{t=1}^{T}$, we have 
\begin{equation}\label{equ_19}
\begin{aligned}
    \mathcal{L}_{dsm} = & \frac{1}{T} \sum_{t=1}^{T} \lambda(\sigma_{t}) \mathbb{E}_{p(x)} \mathbb{E}_{x_{t} \sim p_{\sigma_{t}}(x_{t} \vert x)} \| s_{\theta}(x_{t}, \sigma_{t}) \\
    &+ \frac{x_{t}-x}{\sigma_{t}} \|_{2}^{2},
\end{aligned}
\end{equation}
where $\lambda(\sigma_{t})$ is a weighting factor.

\begin{table*}
 \caption{The training and sampling procedures of our proposed conditional DDPM.}\label{table1}
  \centering
  \begin{tabular}{p{0.45\textwidth}p{0.45\textwidth}}
    \hline
    \textbf{Algorithm 1} Training    & \textbf{Algorithm 2} Sampling \\
    \hline
    1:  \textbf{repeat} & \textbf{Require:} N: Number of steps \\
    2:  \hspace{4mm}$(x_{0}^{i}, y^{i}) \sim p(x,y)$ &1:  $x_{T} \sim \mathcal{N}(0, \mathbf{I})$ \\
    3:  \hspace{4mm}$t \sim \mathcal{U}(\{1,\ldots, T\})$ & 2:  \textbf{for} $i=N,\ldots,1$ \textbf{do} \\
    4:  \hspace{4mm}$\epsilon \sim \mathcal{N}(0,I)$ & 3:  \hspace{4mm}$z \sim \mathcal{N}(0, \mathbf{I})$ if $i>1$, else $z=0$ \\
    5:  \hspace{4mm}Take gradient descent step on & 4:  \hspace{4mm}$x_{i-1} = \frac{1}{\sqrt{\alpha_{i}}}(x_{i} - \frac{1-\alpha_{i}}{\sqrt{1-\bar{\alpha}_{i}}}\epsilon_{\theta}(y,x_{i},t)) + \sigma_{i}z$ \\
    \hspace{7mm} $\nabla_{\theta}\| \epsilon_{t} - \epsilon_{\theta}(y^{i}, \sqrt{\bar{\alpha}_{t}}x_{0}^{i} + \sqrt{1-\bar{\alpha}_{t}}\epsilon_{t}, t) \|^{2}$ & 5:  \textbf{end for} \\
    6:  \textbf{until} converged & 6:  \textbf{return} $x_{0}$ \\
    \hline
  \end{tabular}
\end{table*}

\subsubsection{Stochastic differential equation}
Similar to DDPM and NCSN, the stochastic differential equation (SDE) framwork gradually transforms the original data distribution into a Gaussian distribution in the forward stage. Unlike the other two methods that split the diffusion process into many discrete steps, the SDE method handles a continuous process. Thus, the SDE method can be seen as a generalization of DDPM and NCSN methods. Let us use $p_{t}(x)$ for the probability density function of $x(t)$, and $p_{st}(x(t)\vert x(s))$ for the transition kernel from $x(s)$ to $x(t)$, where $0\leqslant s < t \leqslant T$. Typically, $p_{T}$ is an unstructured prior distribution without information from $p_{0}$. To calculate the diffusion process of the SDE method, it is necessary to solve the following SDE
\begin{equation}\label{equ_20}
dx = f(x, t)\cdot dt + g(t)\cdot d\omega,
\end{equation}
where $t \sim \mathcal{U}([0, T])$, $\omega$ is the Brownian motion, $f$ and $g $ are drift and diffusion coefficients respectively. Similarly, we have the associated reverse-time SDE:
\begin{equation}\label{equ_21}
dx = [f(x, t) + g(t)^{2} \cdot \nabla_{x} \log p_{t}(x)]dt + g(t)d\hat{\omega},
\end{equation}
where $\nabla_{x} \log p_{t}(x)$ is the score function of the data distribution $p_{t}(x)$, $\hat{\omega}$ represents the Brownian motion when time is reversed, and $t \sim \mathcal{U}([T, 0])$. A neural network $s_{\theta}$ is trained to estimate the score so that $s_{\theta}(x, t) \approx \nabla_{x} \log p_{t}(x)$. The objective function is the continuous version of~\eqref{equ_19} and can be expressed as
\begin{equation}\label{equ_22}
\begin{aligned}
    \mathcal{L}_{dsm}^{\ast} = & \mathbb{E}_{t} [\lambda(t) \mathbb{E}_{x(0)} \mathbb{E}_{x(t) \vert x(0)} \| s_{\theta}(x_{t}, t) \\
    &- \nabla_{x(t)} \log p_{0t}(x(t) \vert x(0)) \|_{2}^{2}],
\end{aligned}
\end{equation}
where $\lambda(t)$ is a positive weighting function, $t \sim \mathcal{U}([0, T])$, $x(0) \sim p_{0}(x)$, and $x(t) \sim p_{0t}(x(t) \vert x(0))$. In~\eqref{equ_22}, $\nabla_{x(t)} \log p_{0t}(x(t) \vert x(0))$ is used to replace the original score, as proposed in~\cite{vincent2011connection}.

\begin{table*}
 \caption{The training and sampling procedures of the proposed conditional SDE.}\label{table2}
  \centering
  \begin{tabular}{p{0.45\textwidth}p{0.45\textwidth}}
    \hline
    \textbf{Algorithm 1} Training    & \textbf{Algorithm 2} Euler-Maruyama Sampling \\
    \hline
    1:  \textbf{repeat} & \textbf{Require:} N: Number of discretization steps \\
    2:  \hspace{4mm}$(x_{0}, y) \sim p(x,y)$ & 1:  $x_{N} \sim \mathcal{N}(0, \sigma(T)\cdot \mathbf{I}) $ \\
    3:  \hspace{4mm}$t \sim \mathcal{U}([0,1])$ & 2:  \textbf{for} $i=N,\ldots, 1$ \textbf{do} \\
    4:  \hspace{4mm}$\epsilon \sim \mathcal{N}(0,\mathbf{I})$ & 3:  \hspace{4mm}$t = \frac{i\cdot T}{N}$ \\
    5:  \hspace{4mm}$x(t) = x(0) + \epsilon \sigma(t) $ & 4:  \hspace{4mm}$z \sim \mathcal{N}(0, \mathbf{I})$ \\
    6:  \hspace{4mm}Take a gradient descent step on &  5:  \hspace{4 mm}$x_{i-1} = x_{i} + \sigma^{2t}s_{\theta}(x_{i}, y, t)\frac{T}{N} + \sigma^{t}\sqrt{\frac{T}{N}} z$ \\
    \hspace{7mm} $\nabla_{\theta}\| s_{\theta}(x(t), y, t) + \epsilon \|^{2}$ &  6:  \textbf{end for} \\
    7: \textbf{until} converged & 7:  \textbf{return} $x_{0}$ \\
    \hline
    \textbf{Algorithm 3} Predictor-Corrector Sampling    & \textbf{Algorithm 4} ODE Sampling \\
    \hline
    \textbf{Require:} N: Number of discretization steps & \textbf{Require:} N: Number of discretization steps \\
    \hspace{12.8mm} M: Number of correction steps & 1: $x_{N} \sim \mathcal{N}(0, \sigma(T)\cdot \mathbf{I}) $ \\
    1:  $x_{N} \sim \mathcal{N}(0, \sigma(T)\cdot \mathbf{I})$ & 2: solve ODE $dx = -\frac{1}{2}\sigma^(2t)s_{\theta}(x(t), y, t)dt$ \\
    2:  \textbf{for} $i=N,\ldots, 1$ \textbf{do} & \ \ \ \ \ \ in $t=\frac{i\cdot T}{N}, i \in \{0,\ldots, N-1\}$ \\
    3:  \hspace{4mm}$t = \frac{i\cdot T}{N}$ & 4: find $x_{0}$ via solving ODE at $t=0$  \\
    4:  \hspace{4mm}$z \sim \mathcal{N}(0, \mathbf{I})$ if $i>1$, else $z=0$ & 5: \textbf{return} $x_{0}$ \\
    5:  \hspace{4mm}$x_{i-1} = x_{i} + \sigma^{2t}s_{\theta}(x_{i}, y, t)\frac{T}{N} + \sigma^{t}\sqrt{\frac{T}{N}} z$ & \\
    6:  \hspace{4mm}\textbf{for} $j=1,\ldots, M$ \textbf{do} & \\
    7:  \hspace{8mm}$z_{j} \sim \mathcal{N}(0, \mathbf{I})$ & \\
    8:  \hspace{8mm}$\gamma = \frac{(r\|z\|_{2})^{2}}{\|s_{\theta}(x_{i}, y, \sigma_{i})\|_{2}^{2}}$ & \\
    9:  \hspace{8mm}$x_{i-1} = x_{i-1} + \frac{\gamma}{2}s_{\theta}(x_{i}, y, \sigma_{i}) + \sqrt{\gamma} z_{j}$ & \\
    10: \hspace{2.5mm}\textbf{end for} & \\
    11: \textbf{end for} & \\   
    12: \textbf{return} $x_{0}$ & \\
    \hline
  \end{tabular}
\end{table*}

\begin{figure*}[!ht]
\centering
\includegraphics[width=\textwidth]{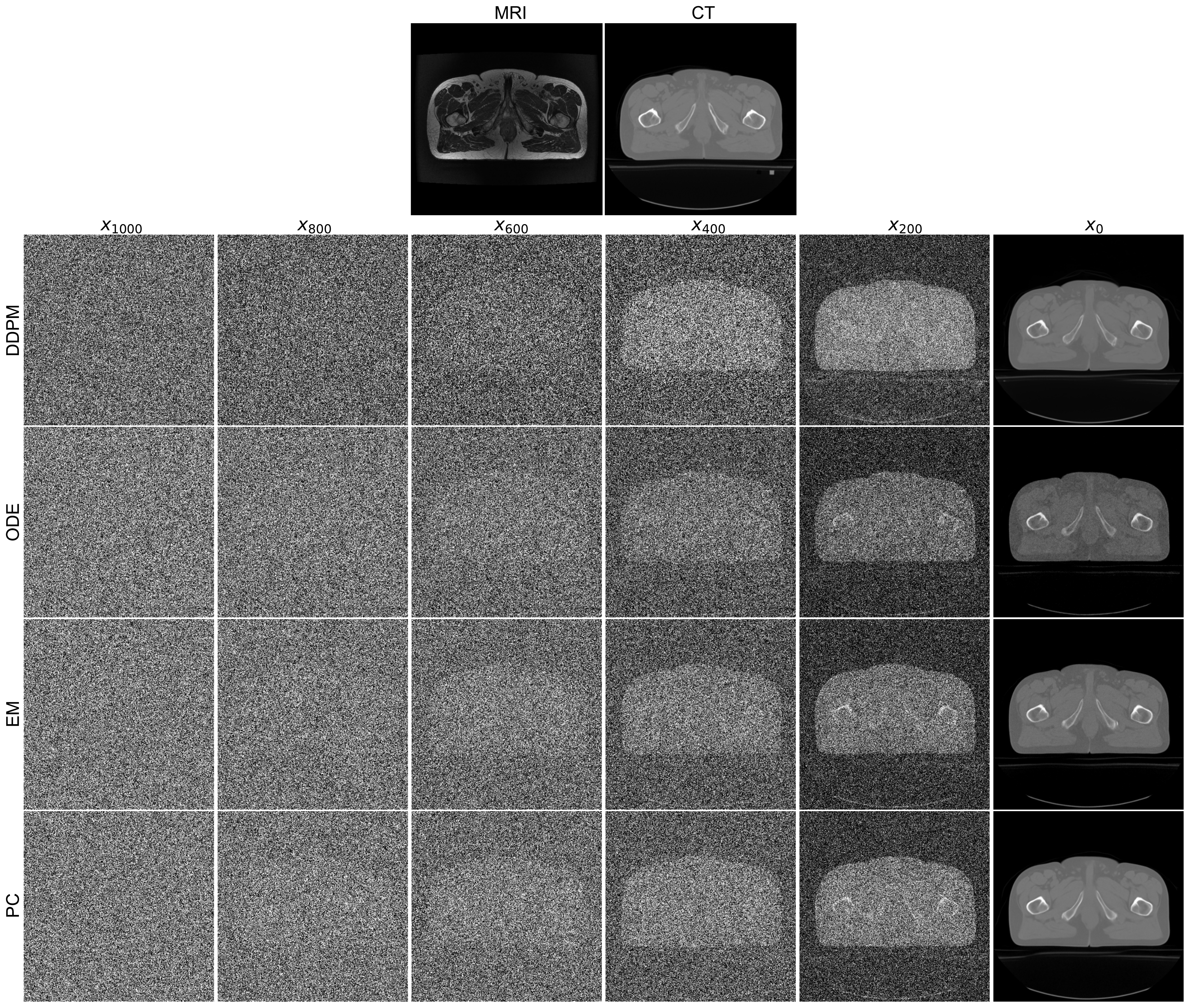}
\caption{Reverse diffusion results using different sampling methods. The first row shows a T2w MR image as the condition and a CT image as the ground truth. The second to fifth rows present intermediate results in the DDPM, ODE, EM, and PC sampling processes respectively.}
\label{fig_1}
\end{figure*}

\section{Methodology}
\subsection{Data}
The Gold Atlas male pelvis dataset~\cite{nyholm2018mr} was utilized in this study, which consists of co-registrated T2w MRI and CT image pairs from 19 patients. Data were collected in three different departments. CT images were obtained on a Siemens Somantom Definition AS+ scanner, a Toshiba Aquilion scanner, and a Siemens Emotion 6 scanner with the pixel size range between $0.98 mm \times 0.98 mm$ and $1 mm \times 1 mm$. T2w MR images were scanned on a GE Discovery 750w scanner with the FRFSE sequence, a Siemens scanner with the TSE sequence, and a GE signa PET/MR scanner with a FRFSE sequence, with the pixel size range between $0.875 mm \times 0.875 mm$ and $1.1 mm \times 1.1 mm$. We randomly selected 17 patients with 1,416 image pairs for training, and the other two patients with 135 image pairs for testing. All image matrices in this study are $512 \time 512$. Before training, all images were pre-processed for pixel intensity unification.

\subsection{Conditional DDPM}
To implement image synthesis between CT and MRI, here we extend the conditional DDPM proposed by Saharia~\etal~\cite{saharia2022image} from working within the same imaging mode (photographs) to mapping across different imaging modes (CT and MRI) so that we can build diffusion and score-matching models conditioned on T2w images. Given the co-registered CT and T2w MRI pairs $(x^{i}, y^{i})_{i=1}^{K}$, where $K$ is the number of image pairs in the dataset, our objective function of~\eqref{equ_14} is as follows:
\begin{equation}\label{equ_23}
\begin{aligned}
    \mathcal{L}_{t}^{simple} = & \mathbb{E}_{t \sim [1,T],x_{0},\epsilon_{t}} [\| \epsilon_{t} - \epsilon_{\theta}(\sqrt{\bar{\alpha}_{t}}x_{0}^{i} \\
    &+ \sqrt{1-\bar{\alpha}_{t}}\epsilon_{t}, y, t) \|^{2}].
\end{aligned}
\end{equation} 
The sampling process is a reverse Markovian process starting from a Gaussian noise $x_{T} \sim \mathcal{N}(0, \mathbf{I})$, the reverse process of~\eqref{equ_6} and~\eqref{equ_9} can be modified as
\begin{equation}\label{equ_24}
q(x_{t-1} \vert x_{t}, x_{0}, y) = \mathcal{N}(x_{t-1}; \tilde{\mu}(x_{t}, x_{0}, y), \tilde{\beta}_{t}\cdot \mathbf{I}),
\end{equation} 
\begin{equation}\label{equ_25}
\tilde{\mu}_{\theta}(x_{t}, y, t) = \frac{1}{\sqrt{\alpha_{t}}}(x_{t} - \frac{1-\alpha_{t}}{\sqrt{1-\bar{\alpha}_{t}}}\epsilon_{\theta}(x_{t}, y, t)).
\end{equation}

\begin{figure*}[t]
\centering
\includegraphics[width=\textwidth]{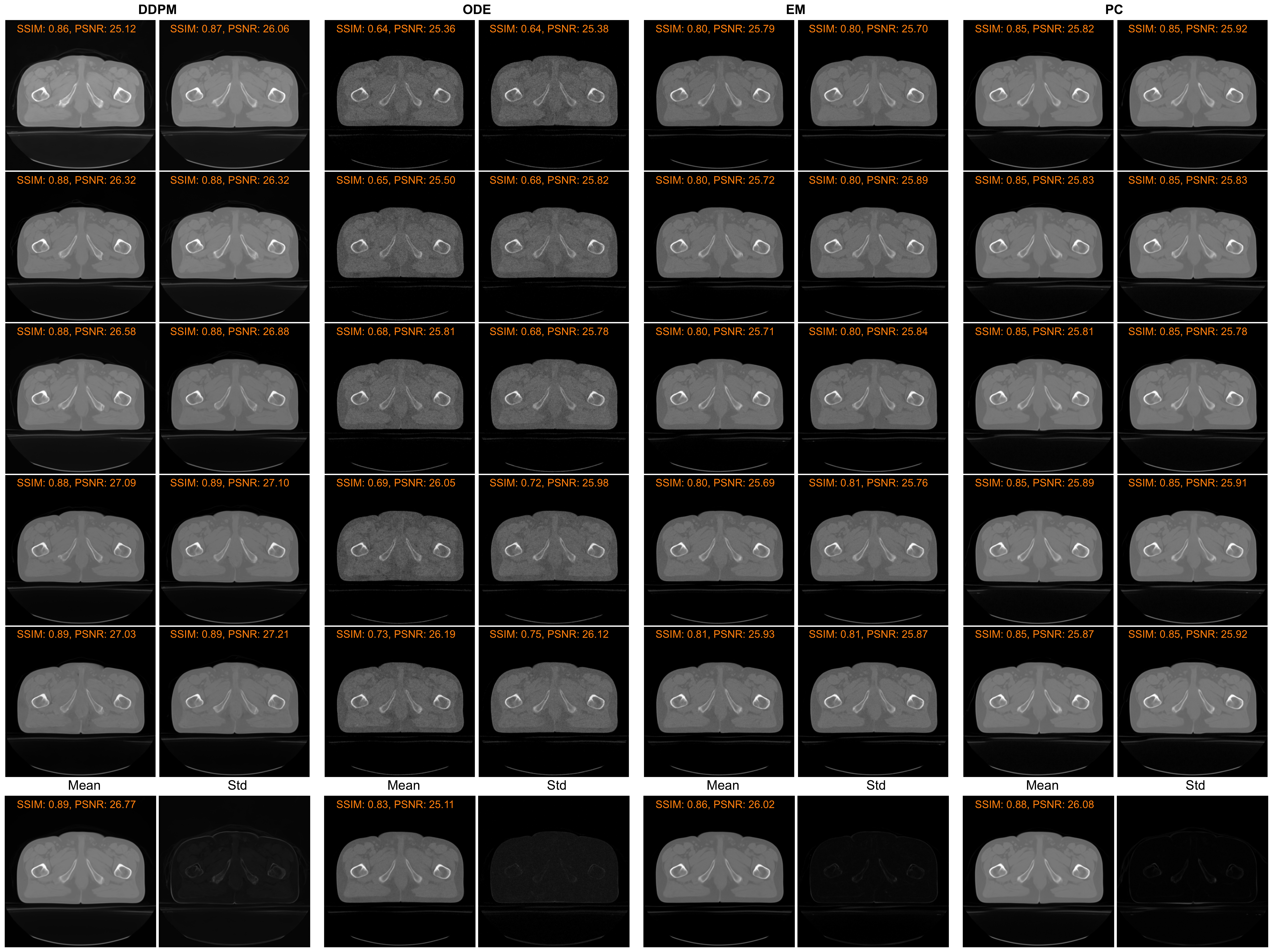}
\caption{Comparison of Monte Carlo sampling results using different reverse methods. The first five rows show the results conditioned on the same MR image using four different sampling strategies respectively. The bottom row presents results averaged over all ten MC sampling results and their pixel-wise standard deviation maps. Notably, all the images are shown in the range of [0, 1] except for the standard deviation maps in the range of [0, 0.5].}
\label{fig_2}
\end{figure*}

The training and sampling procedures of modified DDPM are listed in Table~\ref{table1}. There were 1,000 diffusion steps and 1,000 sampling steps for the modified DDPM. UNet~\cite{ronneberger2015u} was adopted for the reverse diffusion process to denoise.

\subsection{Conditional SDE}
To generate required sampling results, the reverse-time SDE should be solved under the guidance of a condition of interest. There are various approaches for enforcing a condition gently or strongly, such as classifier-free and classier-guidance methods. The classifier-free approach, as mentioned by Song~\etal~\cite{song2020score}, adds a condition in the diffusion model training process and performs the network training in a supervised fashion. Different from the classifer-free approach, the classifier-guidance approach is unsupervised. Song~\etal~\cite{song2021solving} trained a network for unconditional score estimation and then incorporated conditional information into the sampling process with a proximal optimization step based on a physical measurement model for medical imaging such as the Radon and Fourier transforms. Dharwal~\etal~\cite{dhariwal2021diffusion} and Liu~\etal~\cite{liu2021more} added a classifier and used its gradients to regulate the reverse diffusion process of a pre-trained unconditional diffusion model.

In this study, we include T2w MRI images as the condition in the training process. In other words, we supervise the forward and backward diffusion processes. Specifically, we adopted the variance exploding (VE) SDE setting described in the paper~\cite{song2020score} with $f=0$ and $g=\sigma^{t}$. Given the original expression of $g(t)=\sqrt{\frac{d[\sigma^{2}(t)]}{dt}}$ and $\sigma(0)=0$, we have $\sigma(t)=\sqrt{\frac{\sigma^{2t}-1}{2\log \sigma}}$. Hence, equation~\eqref{equ_20} can be changed to the following form:
\begin{equation}\label{equ_26}
dx = \sigma^{t}\cdot d\omega,
\end{equation}
\begin{equation}\label{equ_27}
p_{0t}(x(t)\vert x(0)) = \mathcal{N}(x(t); x(0), \sigma(t)\cdot \mathbf{I}),
\end{equation}
where $t \sim \mathcal{U}([0, T])$. As $p_{0t}(x(t) \vert x(0))$ is a Gaussian perturbation kernel, the gradient of the perturbation kernel is $\nabla_{x(t)} p_{0t}(x(t) \vert x(0)) = - \frac{x(t)-x(0)}{\sigma(t)}$. Accordingly, the objective function becomes
\begin{equation}\label{equ_28}
\begin{aligned}
    \mathcal{L}_{dsm}^{\ast} = & \mathbb{E}_{t}[\lambda(t) \mathbb{E}_{x(0)} \mathbb{E}_{(x(t) \vert x(0))} \| s_{\theta}(x(t), y, t) \\
    &+ \frac{x(t)-x(0)}{\sigma(t)} \|_{2}^{2}].
\end{aligned}
\end{equation}

The reverse-time SDE of~\eqref{equ_21} can be expressed as
\begin{equation}\label{equ_29}
dx = -\sigma^{2t}s_{\theta}(x(t), y, t)dt + \sigma^{t}d\bar{\omega}.
\end{equation}
To sample from the time-dependent score-based model $s_{\theta}(x(t), y, t)$, we first draw a sample $x_{T}$ from the prior distribution $p_{T} \sim \mathcal{N}(x(0), \sigma(T)\cdot \mathbf{I})$, and then solve the reverse-time SDE numerically. When $\sigma(T)$ is large, the mean value of the prior distribution is close to 0, and we can approximate the prior distribution to $p_{T} \sim \mathcal{N}(0, \sigma(T)\cdot \mathbf{I})$. In this study, three sampling techniques were used, which are Euler-Maruyama (EM), Prediction-Corrector (PC), and probability flow ordinary differential equation (ODE) methods respectively.

\subsubsection{Euler-Maruyama method}
In the EM method, to solve the reverse-time SDE of~\eqref{equ_29}, a simple discretization strategy is adopted, replacing $dt$ with a small increment $\Delta t$ and $d\bar{w}$ with a Gaussian noise $z\sim \mathcal{N}(0, \Delta t \cdot \mathbf{I})$. Then, we have
\begin{equation}\label{equ_30}
x_{t-\Delta t} = x_{t} + \sigma^{2t}s_{\theta}(x(t), y, t)\Delta t + \sigma^{t}\sqrt{\Delta t} z_{t},
\end{equation}
where $z_{t} \sim \mathcal{N}(0, \mathbf{I})$.

\subsubsection{Prediction-Correction method}
The PC sampling alternates between prediction and correction steps. The predictor can be any numerical solver for the reverse-time SDE with a fixed discretization strategy, such as the EM method of~\eqref{equ_30}. The corrector can be any score-based Markov Chain Monte Carlo method, such as annealed Langevin dynamics. To implement annealed Langevin dynamics, it is necessary to calculate a Langevin step size $\gamma$:
\begin{equation}\label{equ_31}
\gamma = \frac{(r\|z\|_{2})^{2}}{\|s_{\theta}(x_{i}, y, \sigma_{i})\|_{2}^{2}},
\end{equation}
where $r$ is a signal-to-noise ratio, and $z\sim \mathcal{N}(0,\mathbf{I})$. Once the Langevin step size $\gamma$ is determined, we can sample according to Langevin dynamics of~\eqref{equ_16}.

\subsubsection{Probability flow ODE method}
We call the probability flow ODE method as ODE in this paper for simplicity. For any SDE in the form of~\eqref{equ_20}, there exists an associated ODE
\begin{equation}\label{equ_32}
dx = [f(x, t) - \frac{1}{2}g(t)^{2}\nabla_{x}\log p_{t}(x)]dt,
\end{equation}
which has the same marginal probability density $p_{t}(t)$ trajectory as that of the SDE. As a result, sampling by solving the reverse-time SDE is equivalent to solving the above ODE in the reverse time direction. Being the same as the EM and PC methods, the ODE sampling process starts from obtaining $x_{T}$ from $p_{T}$. Then, we integrate ODE in the reverse time direction and finally get a sample from $p_{0}$. In this case, the ODE equation is written as follows:
\begin{equation}\label{equ_32}
dx = -\frac{1}{2}\sigma^{2t}s_{\theta}(x(t), y, t)dt.
\end{equation}
We used the Explicit Runge-Kutta method of order 5(4) method to solve~\eqref{equ_32}.

The training and sampling procedures of the three mentioned methods are listed in Table~\ref{table2}. We used UNet for score estimation. For all the compared sampling methods, we set the total sampling steps to 1,000. In particular, the PC sampling was set with 500 prediction steps and 500 correction steps.

\begin{figure*}[t]
\centering
\includegraphics[width=6in]{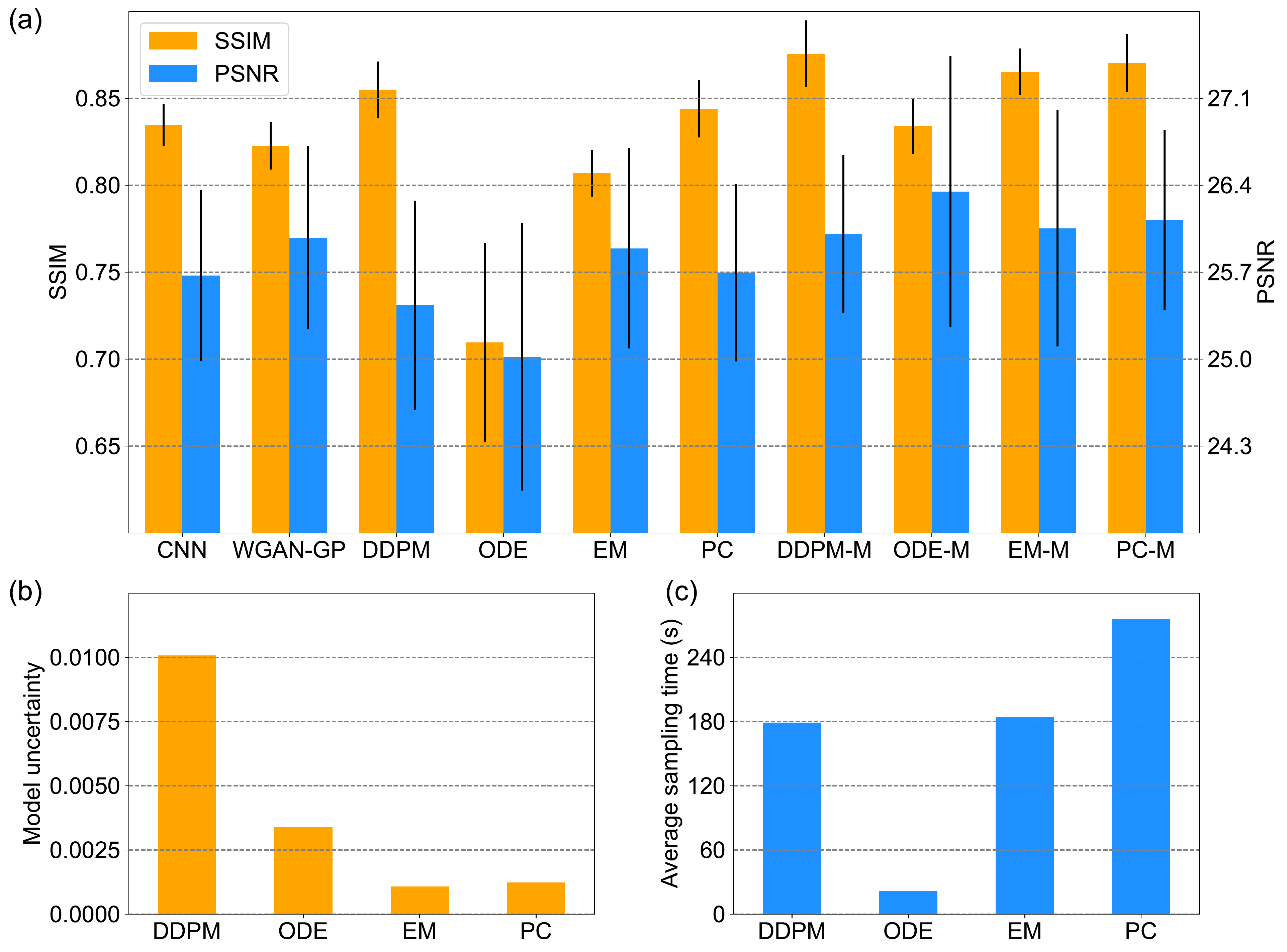}
\caption{Statistical comparison. (a) The average SSIM and PSNR scores of CT images synthesized using different methods, where the error bars show standard deviations; (b) the average model uncertainties of different sampling methods; and (c) the average times for sampling a slice using different sampling methods respectively.}
\label{fig_3}
\end{figure*}

\subsection{Other methods for comparison}
We also compared the diffusion and score-matching models with CNN and GAN based models. For CNN, UNet was trained to minimize MSE. For GAN, Wasserstein distance and gradient penalty were incorporated (WGAN-GP) ~\cite{gulrajani2017improved} with UNet as the generator and MSE as data fidelity measure.

\subsection{Implementation details}
In our experiments, for all the methods including conditional DDPM, conditional SDE, CNN, and WGAN-GP, we consistently used UNet of the same architecture (UNet of CNN and WGAN-GP without time-embedding) and the Adam optimizer with a learning rate of $10^{-4}$, betas of 0.9 and 0.999, and eps of $10^{-8}$. The training process continued at least 100 epochs and stopped after the loss did not decrease by 1\% relative to the average loss for the twenty previous epochs. The batch size was fixed at 2. All the experiments were conducted using PyTorch 1.11 on a 24GB Nvidia RTX Titan GPU. We will upload all the codes for this study onto Github after the publication of this paper.

For the conditional DDPM and conditional SDE methods, since noises were involved in the sampling processes, the final outputs were subject to random fluctuations. Hence, we further investigated uncertainties of the diffusion and score-matching models using the Monte Carlo (MC) method. Each target CT image was generated ten times. In each case, we recorded all sampling results and obtained a MC result by averaging all ten results. We denote these averaged results using DDPM, ODE, EM, and ODE sampling methods as DDPM-M, ODE-M, EM-M and PC-M respectively. The model uncertainty was revealed by looking at the standard deviation map of the ten sampling results in each configuration.

\section{Results}
In this study, we utilized the structural similarity index measure (SSIM) and peak signal-to-noise ratio (PSNR) metrics to evaluate image quality.

\subsection{Diffusion and score-matching results}
Intermediate results from DDPM, EM, PC, and ODE methods in the reverse process are compared in Fig.~\ref{fig_1}. It can be found that all the four methods finally generate a desirable CT image ($x_{0}$) starting from a Gaussian noise ($x_{1,000}$) conditioned on the associated T2w MR image. As the reverse diffusion process goes on, noise is gradually removed to make an image increasingly realistic.

\subsection{Model uncertainty estimation}
Fig.~\ref{fig_2} shows MC results using different sampling methods respectively. It seems that DDPM generates results with the highest SSIM and PSNR scores while the ODE method produces results with the lowest scores. For those results obtained by averaging all ten MC samples conditioned on the same T2w MR image, ODE, EM, and PC methods offer significantly better synthetic CT images with higher SSIM and PSNR scores than the corresponding individual result. In terms of the standard deviation map, the PC and EM methods generate lower standard deviations than the DDPM and ODE methods. Quantitative results in Fig.~\ref{fig_3}(b) shows model uncertainty scores for the sampling methods, which were computed by averaging over a standard deviation map, demonstrating that EM and PC have lower of model uncertainty scores than DDPM and ODE.

\begin{figure*}[t]
\centering
\includegraphics[width=\textwidth]{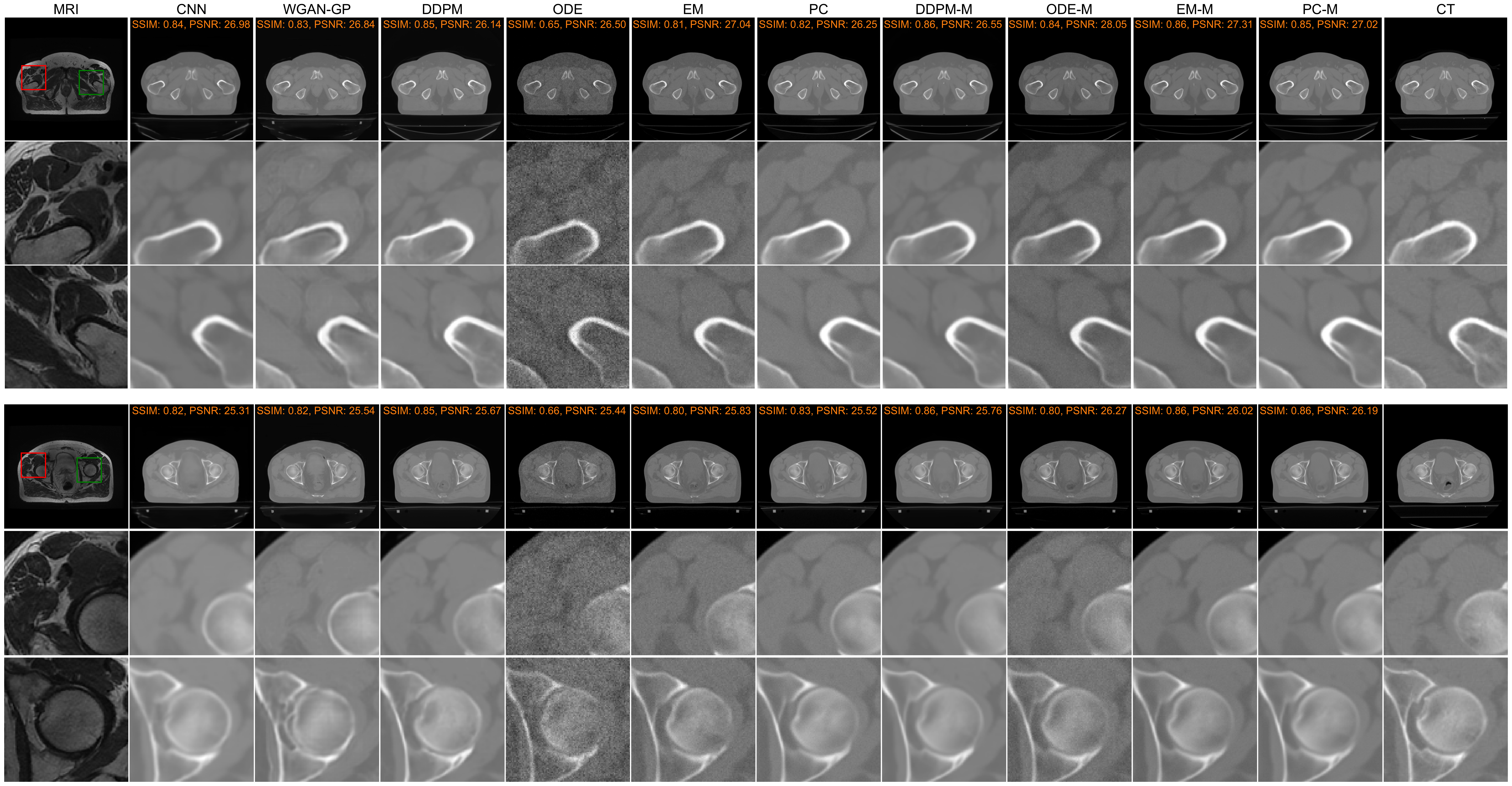}
\caption{Comparison of two image synthesis results using different methods. For each example, the first row shows the whole image, and the second and third rows present the zoomed regions bounded by red and green boxes, respectively.}
\label{fig_4}
\end{figure*}

\subsection{Comparison with CNN and GAN}
We compared diffusion and score-matching models with CNN and WGAN-GP based models on image synthesis between CT and MRI. Qualitative results are shown in Fig.~\ref{fig_4}. It is found that the CNN method tends to generate over-smoothed results. Although the WGAN-GP method generates results with details, it tends to generate artifacts. In the bottom row of Fig.~\ref{fig_4}, the left side of the round bone structure produced by WGAN-GP contains severe artifacts. On the other hand, the diffusion and score-matching results have faithful details. Among the diffusion and score-matching results, the ODE method generates less favorable results than the other three sampling methods, and DDPM and PC methods show higher SSIM scores than the other two sampling methods. Looking at the averaged results (DDPM-M, ODE-M, EM-M, and PC-M), it is found that all four kinds of averaged results are better than the corresponding individual sampling results in terms of information fidelity, SSIM and PSNR scores. Quantitative results in Fig.~\ref{fig_3}(a) also demonstrate higher scores of DDPM-M, ODE-M, EM-M, and PC-M in terms of SSIM and PSNR than DDPM, ODE, EM, and PC respectively. We also investigated inference speed of each sampling method. Fig.~\ref{fig_3}(c) compares the amount of time to generate a $512\times512$ synthetic CT image using each sampling method. The ODE method is the fastest among all the four methods. The sampling time of DDPM and EM are comparable while the PC method is the slowest.

\section{Discussions and conclusion}

In this study, DDPM is time-discrete in both training and sampling processes. In contrast, SDE is time-continuous in the training process and time-discrete in the sampling process. We investigated both DDPM and SDE methods generating synthetic CT images from given T2w MRI. Four different sampling methods (one DDPM based and three SDE based) were compared. According to our results, all the four sampling methods can remove noises and generate realistic CT images. After averaging multiple Monte Carlo sampling results, excellent results can be obtained for all the four methods. In terms of sampling quality, the ODE method brings about inferior results while the other three methods produce comparable good results. However, in terms of sampling speed, the ODE method is significantly faster than the other methods while the PC method is the slowest. Practically, it is necessary to balance between sampling quality and sampling speed in an application-specific manner. Given the good sampling quality and a relative fast sampling speed of the EM method, we would recommend it as a good choice for SDE-based sampling. 

When looking at the model uncertainty in Fig.~\ref{fig_3}(b) and standard deviation maps in Fig.~\ref{fig_2}, DDPM has a greater model uncertainty than the SDE-based sampling methods, we infer that the time-discrete training could be a major reason for the uncertainty of the DDPM model. We further assume that increasing the number of steps in the DDPM diffusion and reverse processes would allow DDPM to perform more like a time-continuous model with a reduced model uncertainty.

We have compared diffusion and score-matching results with CNN-based and GAN-based results. Among all the results, CNN results tend to be over-smoothed, which is partly due to the utilization of MSE as the objective function~\cite{johnson2016perceptual}. GAN results have more details than the CNN results but are compromised by artifacts. These artifacts may come from a low robustness of the GAN model and a high tendency of hallucination. The diffusion and score-matching models, different from CNN and GAN models, have principled abilities to fit data distributions and generate high quality images. However, diffusion and score-matching models suffer from a well-known disadvantage: relying on a long Markov chain to generate results in a relatively slow speed.

As a future direction, we will keep exploring computational techniques to  boost the sampling processes significantly. For that purpose, some algorithms were already proposed~\cite{song2020denoising, lu2022dpm, ma2022accelerating, lyu2022accelerating, chung2022come}, but there is still a large gap between the speed of diffusion sampling and that of CNN and GAN inference.

Compared with the supervised approach, the unsupervised approach does not rely on paired images, and is more useful in medical imaging applications. However, the existing unsupervised diffusion and score-matching models have limitations. The constrains in the methods by Dharwal and Liu~\cite{dhariwal2021diffusion, liu2021more} are too weak to reach specific structures or contents in generated images. On the other hand, the conditions in Song's method~\cite{song2021solving} are too strong, and it is hard to be implemented for CT-MRI synthesis as the similarity between CT and MRI data is not high. In future studies, we will explore opportunities to balance these conditional constrains for high quality CT-MRI synthesis.

In conclusion, we have adapted the emerging diffusion and score-matching models for image synthesis between CT and MRI. The four strategies, including DDPM, ODE, EM, and PC, have been adopted to sample CT images conditioned on an MRI image. The resultant CT images using different sampling strategies have been favorably compared with the results generated using conventional CNN and GAN models. The uncertainties of the diffusion and score-matching models have been quantified as well. Further investigation based on this paper is in progress.

\bibliographystyle{ieeetr}  
\bibliography{references}



\end{document}